\documentclass{elsarticle}

\usepackage[activeacute,english]{babel}
\usepackage[utf8]{inputenc}
\usepackage{subfigure}
\usepackage{graphicx}
\usepackage{amsmath}
\usepackage{color}
\usepackage{multirow}
\usepackage{makecell}
\usepackage{caption}
\usepackage{setspace}
\usepackage{xspace}

\renewcommand*{\k}[0]{\ensuremath{K}\xspace}
\newcommand*{\kk}[0]{\ensuremath{\k^2}\xspace}

\newcommand*{\ktree}[0]{\ensuremath{\k^2}-tree\xspace}
\newcommand*{\ktrees}[0]{\ensuremath{\k^2}-trees\xspace}

\newcommand{\iktree}{I\ktree}

\begin{document}

\begin{frontmatter}

\title{A succinct data structure for self-indexing ternary relations\tnoteref{t1}}
\tnotetext[t1]{\footnotesize
Funded in part by European Union’s Horizon 2020 research and innovation programme under the Marie Sklodowska-Curie grant agreement No 690941 (Project BIRDS).
Sandra Alvarez-Garcia and Nieves Brisaboa were partially funded by MINECO Grant TIN2013-46238-C4-3-R and Xunta de Galicia Grant GRC2013/053. 
Gonzalo Navarro was partially funded by Millennium Nucleus Information and Coordination in Networks ICM/FIC RC130003 and Fondecyt Grant 1-140796.
A preliminary partial version of this article appeared in Proc. DCC 2014, pp. 342--351
}

\author[sag]{Sandra Alvarez-Garcia}
\ead{salvarezg@udc.es}
\author[gdb]{Guillermo de Bernardo\corref{cor1}}
\ead{gdebernardo@udc.es}
\author[nb]{Nieves R. Brisaboa}
\ead{brisaboa@udc.es}
\author[gn]{Gonzalo Navarro}
\ead{gnavarro@dcc.uchile.cl}

\address[sag]{Databases Lab. Campus de Elvi\~na, A Coru\~na, Spain}
\address[gdb]{Enxenio SL. Ba\~nos de Arteixo, A Coru\~na, Spain}
\address[nb]{University of A Coruña. Campus de Elvi\~na, A Coru\~na, Spain}
\address[gn]{DCC, University of Chile, Beauchef 851, Santiago, Chile}

\cortext[cor1]{Corresponding author}

\begin{abstract}

%

The representation of binary relations has been intensively
studied and many different theoretical and practical representations
have been proposed to answer the usual queries in multiple domains.
However, ternary relations have not received as much attention,
even though many real-world applications require the processing of ternary
relations.

In this paper we present a new compressed and self-indexed data
structure that we call Interleaved \ktree (\iktree), designed to
compactly represent and efficiently query general ternary relations.
The \iktree is an evolution of an existing data structure, the
\ktree~\cite{Brisa09}, initially designed to represent Web graphs
and later applied to other domains. The \iktree is able to extend
the \ktree to represent a ternary relation, based on the idea of
decomposing it into a collection of binary relations but providing
indexing capabilities in all the three dimensions. We present
different ways to use \iktree to model different types of ternary
relations using as reference two typical domains:  RDF and Temporal
Graphs. We also experimentally evaluate our representations
comparing them in space usage and performance with other solutions of
the state of the art.

\end{abstract}

\begin{keyword}Compressed Data Structures\sep Ternary Relations\sep RDF\sep Temporal Graphs\sep $K^2$-tree
\end{keyword}

\end{frontmatter}

\section{Introduction}

Graphs are a natural way to represent data. A way to model simple
graphs is to see them as binary relations between two data sets,
taking advantage of the good number of theoretical and practical
results the research about binary relations has produced. In fact,
the efficient representation of binary relations has been
extensively studied and, at present, many theoretical and
practical representations of binary relations have been proposed to
answer the usual queries in multiple domains. Theoretical
representations provide optimal solutions for a large number of
operations and many practical data structures are in use~\cite{binrel1,librogonzalo}. General
binary relation representations are used everywhere: to represent
text, binary matrices,  Web Graphs and, in general, simple graphs.

Ternary relations are also very common, but they have not received
so much attention. Many real-world data can be considered as a
ternary relation and \mbox{handled} using a ternary relation
representation. In addition to pure 3-dimensional data, many binary
relations become ternary relations when other dimension of the
data (usually time) is considered, for example a Web Graph is a
snapshot of the Web pages as binary relations at a specific time
instant, but the evolution of the Web graph over time makes
the relationship among web pages three-dimensional. In the same way,
in general, any collection of 2-dimensional representations where
each one captures a different version of the same data (or an
instant of its temporal evolution), can be seen as a ternary
relation. Among those collections we can point out evolving web
graphs, evolving social networks, etc.

Ternary relations can be also found in other cases, for example, a
bi-dimen\-sional matrix of values can be modeled as a ternary relation
with the values as a third dimension; a simple graph with labels in
its edges is also a ternary relation with the labels as a third
dimension; digital images can also be seem as binary relations among
rows and columns of pixels and the color of each pixel; etc. These
representations arise in many areas:  general raster data, images,
time-evolving raster representing oil patches or in general moving
regions can therefore be seen as a ternary relation.

An interesting example of ternary relations are the RDF graphs used
on the semantic web to represent knowledge. RDF databases are
ternary relations because they are collections of 3-tuples
composed by the values for subject, object and predicate.

In spite of the multiple applications, not much effort has been put
to the \mbox{efficient} representation of general ternary
relations~\cite{barbay2012}. In most of the cases, data that could
be represented and managed as a general ternary relation is managed
using domain-oriented representations and, therefore, general
representations and operation sets are usually overlooked. Examples
of domain-specific representations include
\emph{RDF-3X}~\cite{rdf3x} for RDF graphs, or image
representations such as \emph{tiff} and
\emph{Geotiff}~\cite{geotiff} for both general images and
raster data.

The absence of general ternary relation representations can also be
related to the variability in domain-specific operations over the
data. Pure 3-dimensional queries are required in many applications,
but, in many cases, there are specific requirements for the
different dimensions. An example of this is the representation of
time-evolving graphs, or in general time-evolving binary relations:
the usual operations on the ``spatial'' dimensions are well-defined,
but in the ``temporal'' dimension of the ternary relation the
operations involved include time-instant and time-interval
constraints with special characteristics related to the time domain.
A similar reasoning can be applied to ternary relations that
correspond to general raster data, or any matrix of values: the
``spatial'' dimension has specific semantics but the constraints and
query capabilities in the third dimension differ from those in the
other two.

A usual strategy for the representation of ternary relations, using
representations for binary relations, is called \emph{vertical
partitioning}~\cite{vertpar}.  The idea is to reduce the ternary
relation to a collection of binary relations, hence reducing the
problem to efficiently storing and querying several binary
relations, one for each value of the \emph{partitioning variable}.
This approach can lead to much simpler solutions, but faces the
challenge of efficiently managing query constraints that involve the
dimension used as partitioning variable, since the partition in
separate binary relations will usually provide no indexing
capabilities on the partitioning dimension and hence poor efficiency
in that kind of queries.

The \ktree is a compressed and self-indexed structure initially
designed for Web graphs~\cite{Brisa09,claude10}. It was later used
in other domains~\cite{amcis,dcc} for the compact representation and
efficient querying of binary relations. Its quadtree-like structure
makes symmetric the operations over the two dimensions. This is a
valuable property that other classical approaches do not provide.
For example, an inverted index will provide fast access to the
relations of the elements of one dimension (direct-neighbors), but
to recover the neighbors of the element of the other dimension
(reverse-neighbors) another inverted index will be needed to avoid
sequential navigation.

A collection of \ktrees has been used as an effective representation
of RDF graphs using a vertical partitioning approach~\cite{amcis}.
The great compression capabilities and efficient queries provided by the
\ktrees made that representation really competitive against other solutions of
the state of the art. In this paper we will show how the \iktree goes
farther providing better performance due to its capability to index
the three dimensions in the same data structure.

Summarizing, we introduce in this paper the Interleaved \ktree
(\iktree). The \iktree is a compressed and self-indexed structure to
represent and query general ternary relations that gathers in a
single tree the three dimensions, providing indexing capabilities
over all of them. The \iktree is inspired by the vertical
partitioning approach, and considers one of the dimensions as a
partitioning dimension that is handled differently, but is able to
provide efficient access to any of the three dimensions. The \iktree
is particularly suited to represent relations where general
``spatial'' constraints are imposed on the two main dimensions and
``range'' or ``interval'' constraints are imposed on the third
dimension. We will show the extended capabilities of the \iktree
when compared with simple approaches based on a collection of
independent \ktrees to represent RDF graphs. We will experimentally
apply  the \iktree to obtain a compressed and self-indexed
representation of real RDF datasets, comparing them with
state-of-the-art solutions and also with previous approaches based
on collections of \ktrees.

Finally we show how the \iktree can be used to represent time
evolving graphs, using the changes between two different time
instants as the base for the representation.
\section{ Related work}

\subsection{ The \ktree}

\begin{figure}
\centering
\includegraphics[width=\textwidth]{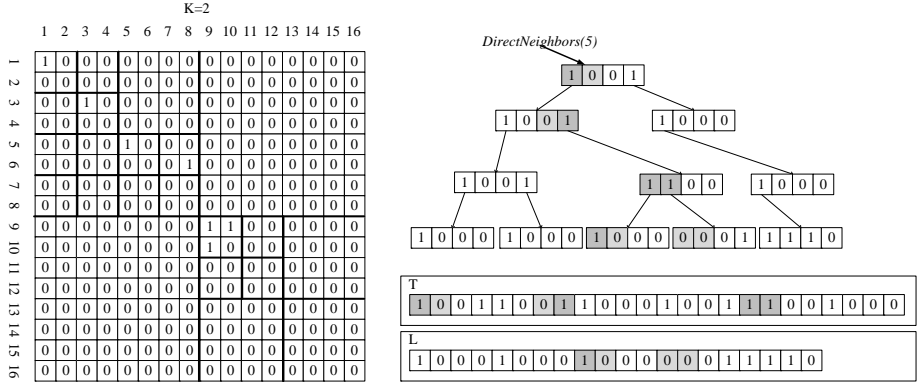}
\caption{An example of a binary relation represented with a \ktree}
\label{fig:k2t}
\end{figure}

The \ktree~\cite{Brisa09} is a compact data structure to represent binary relations,
conceptually represented by a binary adjacency matrix $M$, where $M[i,j]$ is 1
if element $i$ is related with element $j$ and $0$ otherwise. It was
originally designed to represent Web graphs. The $K^2$-tree takes advantage of
the  sparsity of the matrix (large areas of $zeros$) and the clustering
(proximity) of the $ones$. It achieves very low space (less than 5 bits per
link) over Web graphs, allowing large graphs to fit in main memory. It also
supports efficient navigation over the compressed graph \cite{Brisa09},
 efficiently answering  direct and reverse neighbor queries, individual cell and
range queries.

The $K^2$-tree conceptually subdivides the
adjacency matrix
into $K^2$ submatrices of equal size. Each
of the $K^2$ submatrices is
represented with one bit in the first level
of the tree, following a left to
right and top to bottom order. The
bit that represents each submatrix will be
$1$ if the submatrix
contains at least one cell with value $1$. Otherwise, if
it is an empty area, the bit will be a 0. The next level of the tree is
created by expanding the $1$ elements of the current level, subdividing the
submatrix they represent. In this way, $K^2$ children are created in the next
level to represent the new subdivisions. This method continues recursively
until the subdivision reaches the cell-level. Fig.~\ref{fig:k2t} shows an
example of this tree  for
$K=2$. The first $1$ of
the first level (root)
means the upper-left $8\times8$ submatrix has at least a cell with value $1$.
The second bit of the root is a 0, which shows that the upper-right submatrix
does not contain any relation between
nodes, and so on.

The
$K^2$-tree is stored with only two bitmaps: $T$ for the intermediate levels of
the  $K^2$-tree, following a levelwise traversal, and \emph{L} for the bits
of the last
level (Fig.~\ref{fig:k2t}).

Retrieving direct or reverse
neighbors requires obtaining the cells
with value $1$ for a given row or
column in the adjacency matrix.
Both operations are symmetric. They are
solved in the
$K^2$-tree by a top-down traversal over
the tree for the
two appropriate branches of each node. The example shows the
bits of
the tree traversed in order to obtain the direct neighbors of row 5 (i.e., the $ones$ in the 5th
matrix row).

This
navigation over the $K^2$-tree is  efficiently performed over the
bitmaps $T$
and $L$. Given a $1$ at position $x$ in $T$, the children of $x$ are $K^2$
bits placed in $T:L$ starting at position $rank_1(T, x)\times K^2$, where $rank_1$
counts the number of ones in $T[1..x]$. Rank operations are performed in constant time by using an additional \emph{rank structure}, created over the bitmap $T$, that requires sublinear space in addition to $T$ \cite{rank}.

In the worst case the space in bits is $K^2 e(\log_{K^2}{\frac{n^2}{e}} + O(1))$, where $n$ is the number of nodes and $e$ the number of $ones$. Retrieving direct or reverse neighbors in the worst case is $O(n)$ time, although the time is much better in practice. 

The implementation of the $K^2$-tree allows using different $K$ values depending on the level of the tree (\textit{hybrid} approach) or compressing the last levels of the conceptual tree using a vocabulary of submatrices encoded with a statistically compressor. \emph{Direct Access Codes }~\cite{DAC13} (DAC) are used to provide direct access to each code. A dynamic variant of the $K^2$-tree  that combines good compression ratios with fast query and update times has been proposed \cite{Gui12}. Other variants compress efficiently not only large regions of zeros but also regions of ones  \cite{BernardoRoca13}.

The \ktree can be generalized to represent datasets of any number of dimensions building a $K^n$-ary tree instead of the original $K^2$-ary, and adapting the traversal algorithms accordingly. This generalization, a $K^3$-tree or $K^n$-tree, has however implicit scalability problems as $n$ increases due to the structure of the conceptual tree. Since each node of the conceptual tree contains always $K^n$ children, and the bits for all of them must be stored even if only a few of them are $ones$, lots of additional space may be necessary to store this information in many nodes. The \ktree is able to provide a very small representation of a binary relation taking advantage of sparsity and clusterization of the $ones$, but in 3-dimensional or generally in $n$-dimensional datasets it is very difficult to find domains where the $ones$ are highly clustered. Because of this, a $K^n$-tree is unfeasible in general $n$-dimensional datasets.

\subsection{Representation of RDF databases}

The Resource Description Framework (RDF~\cite{manola2004}) is a standard for the representation of information. It models information as a set of triples $(S,P,O)$ where $S$ (subject) is the resource being described, $P$ (predicate) is a property of the resource and $O$ (object) is the value of the property for the subject. RDF was originally conceived as a basis for the representation of information or metadata about documents. 

RDF provides a framework for the conceptual representation of information. As we said, it represents the information as a set of triples $(S,P,O)$. These triples can also be seen as edges in a labeled directed graph. The vision of a set of RDF triples as a graph is called RDF graph in the original recommendation~\cite{manola2004}.

RDF datasets can be queried using a standard language called SPARQL~\cite{SPARQL}. This language is based on the concept of \emph~{triple patterns}: a triple pattern is a triple where any of its components can be unknown. In SPARQL this is indicated by prepending the corresponding part with \texttt{?}. Different triple patterns are created simply changing the parts of the triple that are variable. The possible triple patterns that can be constructed are: $(S,P,O)$, $(S,P,?O)$, $(?S,P,O)$, $(S,?P,O)$, $(?S,P,?O)$, $(S,?P,?O)$, $(?S,?P,O)$ and $(?S,?P,?O)$. For instance, $(S,P,?O)$ is a triple pattern matching with all the triples with subject $S$ and predicate $P$, therefore it would return the values of property $P$ for the resource $S$. $(S,?P,?O)$ is a similar query but it contains an \emph{unbounded predicate}: the results of this query would include all the values of \emph{any} property $P$ of subject $S$. 

SPARQL is a complex language, similar to the SQL of relational databases, and it supports multiple selection clauses, ordering and grouping, but its main features are based on the triple-pattern matching and join operations, that involve merging the results of two triple patterns. For example, $(?S,P_1,O_1) \bowtie (?S,P_2,O_2)$ is a join operation between two triple patterns where the common element $?S$ is the join variable. The result of this join operation would contain the resources $S$ whose value for property $P_1$ is $O_1$ and their value for $P_2$ is $O_2$. 

\subsubsection{Alternatives for the representation of RDF graphs}

RDF is only a conceptual framework that does not enforce any physical representation of the data. The recent popularity of RDF has led to the appearance of many different proposals for the actual storage of information in RDF, known as \emph{RDF stores}. Some approaches to represent RDF triples are based on relational databases~\cite{Sakr}. Nevertheless, multi-indexing native solutions are more frequently used~\cite{rdf3x,Weiss2008}. 

A technique usual in RDF is vertical partitioning~\cite{vertpar}. In this approach, since the number of predicates in an RDF datasets usually has a moderate size, the set of predicates is used as a ``partitioning variable'' and the complete RDF dataset is partitioned into a collection of bi-dimensional datasets, each containing the triples associated with one of the predicates. Following this philosophy, an approach for RDF based on a collection of \ktrees was
already studied and shown to be competitive with the state of the art~\cite{amcis}. 
 
\section{Our proposal: the \iktree}\label{sec:proposal}

Consider a ternary relation $Y$ defined as a set of triples $\{(x_i,y_j,z_k)\} \subseteq X \times Y \times Z$. Our structure, called Interleaved \ktree or \iktree, is designed to represent the relation in a single structure, but following a vertical partitioning approach where one of the dimensions will be considered a ``partitioning dimension'' or ``partitioning variable'' and will be managed in a different way. Usually the smaller dimension, that is, the one with less possible values, will be considered as the partitioning dimension, but additional considerations on the required operations may be taken into account. We will consider in our examples that $Y$ is the partitioning dimension. The \iktree starts from the representation of the ternary relation as a collection of $|Y|$ binary relations, one for each different value $y_j \in Y$. We may consider an adjacency matrix representation of the binary relation, where rows represent the values of the variable $X$, while columns represent the values of the variable $Z$. A one in a cell $(x_i,z_k)$ of the adjacency matrix $y_j$ implies the existence of the triple $(x_i,y_j,z_k)$.

Each adjacency matrix could be stored through an independent \ktree in order to compose a complete system containing the full ternary relation. This would lead to a simple vertical partitioning approach with no indexing capabilities over the partitioning dimension. The \iktree is able to combine  this vertical partitioning philosophy with a representation that conceptually divides the ternary relation in several binary relations as explained, but gathers all of them in a single tree structure, providing indexing capabilities in the three dimensions. 

\subsection{Conceptual data structure}

The Interleaved \ktree can be considered as a combination of a collection of \ktrees into a single tree structure that is able to group the information of all the trees, providing efficient access to multiple \ktrees at once and at the same time efficient and simple navigation algorithms similar to those of a single \ktree. The \iktree will be a \kk-ary tree, where each node of the new tree will contain information about all the different values of the partitioning variable. In fact, the \iktree will contain in each node one bit for each of the individual \ktrees that would have a valid node in that path. This means that the \iktree can be seen as a reorganization of the bits of a collection of \ktrees in a single structure, following a specific arrangement so that the structure can be efficiently navigated and accessed.

The \kk nodes in the first level of the \iktree will always contain $|Y|$ bits each, one per each different value $y_j \in Y$. Inside a node, the $j$-th bit will be a 1 if the submatrix that is represented by that node has at least a $1$ for the $y_j$ value. Otherwise, this position will contain a $0$. In other words, the $j$-th bit in a node will be a 1 iff the \ktree that would represent the values for $y_j$ would contain a 1 in the node at the equivalent path.

Each node of the \iktree that contains at least a 1 in its associated bitmap will have exactly \kk children nodes. The number of bits in each child is given by the number of bits with value $1$ in its parent. Hence, for a node $N_i$ with $m$ $ones$, each one of its \kk children will contain $m$ bits, one per value $y_j$ with at least a $one$ in the matrix \mbox{corresponding} to $N_i$.

\begin{figure}
  \centering
      	\includegraphics[width=\textwidth]{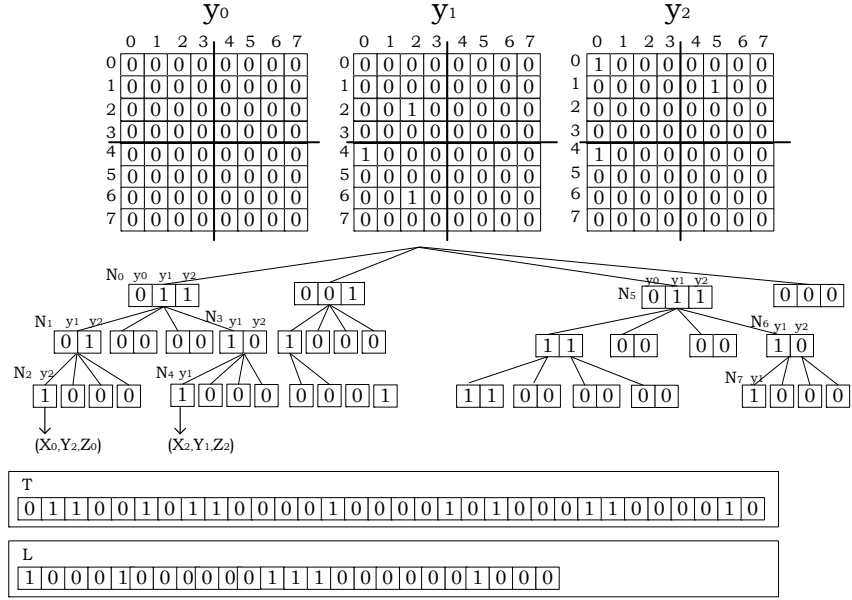}
  \caption{A ternary relation represented with the Interleaved \ktree}
  \label{fig:IKStructure}
\end{figure}

Figure \ref{fig:IKStructure} shows an example of the \iktree structure, representing a simple ternary relation. The ternary relation is already represented  after a vertical partitioning is applied to the smaller dimension $Y$, that contains only 3 possible values and is used as partitioning variable. The adjacency matrices for each value of $Y$ are shown on the top of the figure. On the bottom the complete \iktree structure is shown. A value $K=2$ was chosen for that example, so 4 root nodes appear in the first level of the tree (these can be seen as 4 child nodes of an omitted root node). Each node in the root of the tree contains three bits, one per value of $Y$. The first bit of the first root node ($N_0$) contains a 0, meaning the top-left submatrix of the adjacency matrix for $y_0$ does not contain any $one$. However, the second bit of $N_0$ is a $one$ because the top-left submatrix of $y_1$ contains at least a $one$. 

Given a node of this tree, its children are built recursively considering only the information on the root node and the information in the corresponding sub-matrix following the quadtree-like subdivision. Assuming we have a root node of the tree of size $|Y|$ with $m$ $ones$, its \kk children will be created, following the matrix subdivision, but each child will contain only $m$ bits, that is, the number of \textit{active} values of the parent. The remaining $|Y|-m$ elements in the root node correspond to values of $Y$ that do not contain any element in the corresponding submatrix; therefore they do not need to be decomposed. Therefore, if all the values in the bitmap of a node are $zero$ then that node will not produce any child. The process continues recursively as in the original \ktree, building the conceptual tree top-down until the leaves are reached. Notice that each element in a node will be a bit, corresponding to a different value of $Y$ in increasing order. This will allow us to keep track of the actual values of $Y$ represented by each bit during the navigation of the tree.

\subsection{Physical representation and basic navigation}

The conceptual \iktree is stored following the same ideas used by original \ktrees. A breadth-first traversal of the tree is performed, reading the bitmaps in each level of the tree from left to right into a final bitmap representation. The last level of the tree will be stored as a bitmap $L$, and the other levels will be stored together in a single bitmap $T$. The reason for the use of 2 bitmaps is the same that in the \ktree: $rank$ operation support is needed in the upper levels of the tree to support navigation to the children of a node, while in the last level it is not necessary.

The \iktree construction process yields simple navigation algorithms over the conceptual tree, following the same ideas than a single \ktree. Although nodes in the \iktree have a variable size (depending on the \textit{active} values in the parent), the navigation over the tree is quite similar. Notice that a $one$ in a node is still producing \kk bits in the lower level, and a $zero$ is producing no bits. This is identical to the \ktree, with the only difference that the bits induced by a node in the next level may not be consecutive now because nodes can have several bits, and the bits of a node will be placed together. 

The original \ktree structure has the interesting property that the position of the first child of a node can be computed directly. Since each $one$ of the tree produces $K^2$ children, and the elements are stored in the bitmap ordered by levels, the first children of a node in the position $i$ in the bitmap is in the position $rank_1(T,i-1)*K^2$ of the bitmap $T:L$. 

In the \iktree, a node is defined by its starting position in the bitmap and the length of its bitmap. This information is trivially encoded in the root nodes, and must be kept during the top-down navigation of the tree. The position of the children of a node in the bitmap can be computed through a simple formula. Consider a node starting at position $i$ with $b$ bits. First we must count the number $m$ of $ones$ it contains. If $m=0$ it has no children. If $m> 0$, its children will have $m$ bits each. Its first child node will start at position $rank_1(T,i-1)*K^2+adjust$, where $adjust=|Y|*K^2$. Trivially, starting at this position, the next $K^2*m$ bits are representing the children of that node ($K^2$ nodes of size $m$, storing the $m$ bits of each child node consecutively). Note that the formula is almost the same as in the original \ktree, except by the adjusting factor $adjust$ accounting for having $|Y|*K^2$ bits in the first level instead of $K^2$ as in the \ktree. Like in a \ktree, when the position of a node exceeds $|T|$ the navigation continues in $L$ as if the bitmaps were consecutive.

\subsubsection{Leaves compression with DAC}

In the original \ktree compression could be improved significantly by collapsing a few of the lower levels of the tree and statistically compressing the resulting small subtrees. This improvement can also be applied to the \iktree with a few adjustments. Note that in the \iktree each node has an arbitrary number of bits, and each $one$ of that node produces $K^2$ nodes in the next level. Hence, it is more difficult to encode a submatrix vocabulary of complete nodes, since a node in the third level (starting from the bottom of the tree) collapses a cube of dimensions $K^2 \times K^2 \times m$ where $m$ is the number of $ones$ of that node. In order to obtain a collection of equally-sized submatrices that can be easily stored in a vocabulary, the $K^2 \times K^2$ submatrix that each $one$ of a node would individually produce is stored. This means that the considered submatrices are a subregion of an individual adjacency matrix. All of these submatrices are used to compute a global submatrix vocabulary, that will be stored as a compressed sequence represented using DAC. Notice also that this slightly modifies the structure of the original proposal, since now in the compressed representation of $L$ all the bits that represent the children of a node in the next-to-last level of the conceptual tree will be placed together (represented as a dictionary entry). In the original these bits were scattered in the bitmap $L$ to allow the $m$ bits of a node to be placed consecutively. 

Note that, taking into account that each individual submatrix corresponds to an specific value of the variable $Y$, those submatrices could be encoded using a different vocabulary for each $y_j$. This would in practice lead to the creation of $Y$ vocabularies, each identical to the vocabulary representation that an individual \ktree for the corresponding value would store. However, in the \iktree usually all the matrices will be encoded with the same vocabulary in order to avoid storing $Y$ independent matrix vocabularies. 

\subsection{Query algorithms}\label{sec:intnav}

The query algorithms supported by the \iktree are based on the basic navigation operations defined earlier, and consist of simple top-down traversals of the tree to retrieve the values in the appropriate nodes of the conceptual tree for each query. The main operations supported by the \iktree involve arbitrary range constraints in each of the dimensions. We will consider our basic operations denoted by a triple pattern $(x,y,z)$, where each item in the triple represents a fixed value, a range constraint or an unbounded value in one of the dimensions. 

Constraints that are applied over the ``regular'' dimensions $X$ and $Z$ are handled using the simple \ktree algorithms for tree traversal: at each step of the navigation, the branches that intersect with the queried region are determined easily thanks to the fixed-size space partitioning used by the \ktree. On the other hand, constraints that involve the partitioning dimension $Y$ are handled in a completely different way, due to the asymmetric representation used in the \iktree. Since the method to solve constraints in the other dimensions is already known, we will  categorize queries according to the constraints imposed to the partitioning dimension. We will distinguish three main kinds of queries: those in which the partitioning variable is fixed to a specific value (``fixed partitioning variable''), queries in which the partitioning variable can take any value (``unbounded partitioning variable'') and queries where the partitioning variable is restricted to a continuous range (``fixed-range partitioning variable''). Each query type will be solved using a slightly different method depending on the type of constraint used.

\subsubsection{Query patterns with fixed partitioning variable}

When the partitioning variable is fixed (a single value) we must perform in the \iktree the equivalent of a query in a single \ktree. This is independent of the filters in the other dimensions, that only determine the branches that are explored. Let us consider the simplest query, the pattern $(x_i,y_j,z_k)$ that searches for an individual triple in the dataset (that is, whether $(x_i,y_j,z_k)$ belongs to the ternary relation). To answer this query, a single branch is explored in each level of the tree. In the corresponding node for each level of the tree, the bit corresponding to the value $y_j$ is checked. If it is a $zero$, the element was not found and the algorithm returns. Otherwise, the children of the current node are located and the navigation continues in the corresponding child node; the node traversed is determined by the constraints in dimenxions $X$ and $Z$. When we navigate to the child, we must keep track of the offset in the bitmap that represents the location of the bit $y_j$: this is computed by counting the number of $ones$ in the parent up to the bit representing $y_j$ in it. This counting operation can be easily computed in constant time using additional rank operations at each basic navigational step of the conceptual tree.

Figure \ref{fig:IKStructure} shows the nodes involved to solve the query $(x_6,y_1,z_0)$. In the root, the node $N_5$ is explored (because it corresponds to $(x_6,z_0)$) and the second bit (corresponding to $y_1$) is checked. It is a $one$ and it is the first one of the node, so in its children the position for $y_1$ is the first one. The corresponding node for $(x_6,z_0)$ is $N_6$ and the first bit is checked. The process continues until the leaves, where the only  bit of $N_7$ is checked and it is a $one$, so the triple $(x_6,y_1,z_0)$ exists in the dataset.

Any other queries involving a fixed value in the partitioning dimension are solved in the same way, applying the \ktree basic navigation techniques to answer queries involving ranges in the other two dimensions. The only modification when a range or an unbounded constraint is applied in the other dimensions is that several branches may need to be checked for each explored node with a $one$ in the value corresponding to $y_j$.  

Even though the algorithms are very similar in the \iktree and the \ktree, notice that queries with fixed $y$ in the \iktree require additional rank operations for the navigation. These operations represent most of the traversal cost of the tree; hence the added operations may make the \iktree significantly slower than a single \ktree in this kind of queries.

\subsubsection{Unbounded partitioning variable} 

The second type of queries involve no constraint in the partitioning variable. Again, we consider the simplest query of this type, the pattern $(x_i,?,z_k)$, where $?$ denotes an unconstrained dimension; hence the pattern matches all the values of $Y$ related to the pair of elements $(x_i,z_k)$. Following the same principles used for fixed-partitioning-dimension queries, all queries involving ranges or unbounded values in the other dimensions can be straightforwardly adapted from this one. The process to answer this query is similar to the previous one, starting at the root and selecting the appropriate node at each level of the tree. Navigation will only stop if the current node is a leaf node, that is, if no values of $Y$ exist for the current node. In the upper levels of the tree, this means that the current node contains no $ones$ in its bitmap. In this case, navigation ends and no results are returned. Otherwise, the child node is recursively traversed until the last level is reached. In each level, a node may be smaller than its parent since only the values of $Y$ that contain a $one$ in the corresponding submatrix will be represented in the children of each node. In order to keep track of the association between bits of the node and values in $Y$, a list of the active values has to be managed and updated in each level. If we name $A_j$ the list of active values in the node $N_j$, in a root node $N_j$, its list $A_j$ will always be $Y$. The list of active values $A_k$ of a child of a node $N_j$, $N_k$, contains $m$ elements, where $m$ is the number of ones in the node $N_j$, where $A_k[i]=A_j[rank(N_j,i)]$ (the $rank$ operation in the node denotes counting the number of ones in the fragment of $T$/$L$ corresponding to $N_j$). In this assignment, the $i$-th position of the children active list always contains the value corresponding to the $i$-th $one$ in the parent node. It is easy to see that, given the active list of the parent ($A_j$) and the node $N_j$ the list of the active values for the children of the nodes can be computed by a sequential checking over the bits of the node. 

For instance, returning to the example of the Figure \ref{fig:IKStructure}, the query $(x_2,?,z_2)$ starts by checking the first node of the tree, $N_0$, with  a list of active values $A_0=\{0,1,2\}$, because it is a root node (so it initially contains all the different values for the variable $Y$). The values for the variables $X$ and $Z$ determines that the explored child is $N_3$. Since in $N_0$ only the second and third bits have value $one$, the list of active values for $N_3$ is $A_3=\{1,2\}$ which determines that the first bit of $N_3$ (which is a $one$) corresponds to $y_1$ and the second bit (a $zero$) corresponds to $y_2$. As $N_3$ continues having at least a $one$ value, the process continues traversing down the tree until the leaves level, where node $N_4$ is explored. The list of active values of the node $N_4$ is computed from the bitmap of $N_3$ and the list of active values $A_3$. In that way, only the first bit of $N_3$ is a $one$, so $A_4=\{A_3[1]\}=\{1\}$. Then, the bit $one$ located in $N_4$ corresponds to $y_1$ and the final result of the query $(x_2,?,z_2)$ is $\{(z_2,y_1,z_2)\}$.

\subsubsection{Fixed-range partitioning variable} 

The third type of query that may be of interest in a ternary relation involves a range of values in the partitioning variable. In the simplest of these queries, we have a pattern $(x_i,[y_{jl}-y_{jr}],z_k)$, that is, a range of possible values in the partitioning variable. These queries can be solved using a combination of the procedures explained for the fixed-value and unbounded case. We need to keep track of the offsets in the bitmap of each node that correspond to the query range, which can be easily computed as in the fixed-value case, requiring in this case two additional rank \mbox{operations} at each node to keep track of the start and end of the range. Additionally, we need the same list of active values $A$ to keep track of the $y_j$ values associated to each bit in the node, in order to properly answer the query. 

Notice that, like in the case of fixed-value queries, the overhead required to perform navigation in the \iktree will make it slower than querying a single \ktree. However, when ranges are involved, queries cannot be answered directly in a single \ktree but would require instead to perform a synchronized traversal of multiple \ktrees in order to return a result. These operations depend on the number of queried \ktrees, and are in general very inefficient for a large number of them. Hence, queries with fixed-range partitioning dimension are expected to be much more efficient in an \iktree than in a simpler approach based on a collection of independent \ktrees as the size of the ranges increases. Queries with unbounded partitioning variable can be considered as a special case of the fixed-range queries, and would be in practice the best-case scenario when comparing the \iktree with a collection of independent \ktrees.

\section{Using the I$K^2$-tree to Represent RDF Databases }\label{sec:apps}

A representation of RDF databases using a collection of \ktrees has already been proposed~\cite{amcis}, and it was experimentally shown to be competitive with the state of the art. In this section, we aim to take advantage of the  increased indexing capabilities of the \iktree in this domain, where many relevant operations involve unbounded-partitioning-variable queries, that is, they include RDF triple patterns with an unbounded predicate.

We build an \iktree representation and the equivalent representation based on multiple \ktrees. We follow a hybrid approach using $K=4$ in the first five levels of the tree and $K=2$ in the rest of the levels; the same values are used always for the \iktree and for the multiple \ktree. The last levels of the trees
(submatrices of size $8 \times 8$) are statistically compressed using DACs
\cite{DAC13}. Notice that the multiple $K^2$-tree approach compresses each matrix using an
independent vocabulary, which can model the distribution of each adjacency
matrix better than the single vocabulary of submatrices $8 \times 8$ used in
the I$K^2$-tree. However, the single vocabulary of the I$K^2$-tree is an advantage if the number of predicates is large because it avoids storing too many small vocabularies.

\subsection{Experimental framework and results}

In our experiments we analyze the results using four datasets corresponding to different domains and with very different characteristics. The Jamendo dataset\footnote{\texttt{dbtune.org/jamendo}} is a repository of music; 
Dblp\footnote{\texttt{dblp.l3s.de/dblp++.php}} stores Computer Science journals and proceedings;
Geonames\footnote{\texttt{download.geonames.org/all-geonames-rdf.zip}} is a geographic database;
and DBpedia\footnote{\texttt{wiki.dbpedia.org/Downloads351}} is an encyclopedic dataset extracted from Wikipedia.
All our tests are run on an AMD-PhenomTM-II X4 955@3.2 GHz, quad-core, 8GBDDR2, running Ubuntu 9.10. The code was developed
in C, and compiled using gcc (version 4.4.1) with full optimizations enabled -O9.

 Table \ref{tab:spatialRDF} shows the size of the different RDF datasets and the
 compression results obtained by the I$K^2$-tree, the multiple
$K^2$-tree (M$K^2$-tree) and RDF-3X. The first columns show the number of triples and the number of predicates that
each dataset contains. The number of predicates determines the number of independent \ktrees that would be used in the multiple \ktree approach, and also the number of bits in each of the root nodes of the \iktree.

\begin{table}
\footnotesize
\centering
\begin{tabular}{|l | r | r | r | r| r |}
\hline
Dataset & $|Triples|$ & $|Predicates|$ & M$K^2$-tree &   I$K^2$-tree & RDF-3X\\
\hline
\hline
Jamendo & 1,049,639 & 28 & 0.74 & 0.74 & 37.73 \\
\hline
Dblp & 46,597,620 & 27 & 82.48 & 84.04 & 1,643.31 \\
\hline
Geonames & 112,235,492 & 26 & 152.20 & 156.01 & 3,584.80 \\
\hline
DBpedia & 232,542,405 & 39,672 & 931.44 & 788.19 & 9,757.58 \\
\hline
\end{tabular}
\caption{Space comparison for different RDF datasets (in MB)}
\label{tab:spatialRDF}
\end{table}

The multiple \ktree approach obtains slightly better space results in the  datasets Jamendo, DBLP and Geonames. However, in the DBPedia dataset the \iktree obtains much better compression. The difference in space is due to the fact that a matrix vocabulary is used to statistically compress the lower \mbox{levels} of the conceptual trees, since the plain bitmaps of the \iktree are strictly identical to those of the individual \ktrees. The \iktree uses a global vocabulary, while the multiple \ktree approach stores a different vocabulary per \ktree; hence, the \iktree saves some redundancy in the DBPedia dataset, where many small vocabularies must be created for a large number of predicates, while in the smaller datasets the specific vocabularies are able to obtain better compression than the global one in the \iktree. Notice that in all the datasets both representations obtain much better compression than RDF-3X.

Regarding query efficiency, in order to compare the relative efficiency of the \iktree with a multiple \ktree approach we test all the basic RDF triple \mbox{patterns}. We built a collection of 500 queries for each query pattern, and measured average query times for each query pattern in each of the representations. Table \ref{tab:tempRDF1} shows the results for the Geonames dataset (as a representative domain with few predicates) and for DBpedia (as an example with many predicates). 

\begin{table}
\centering
\begin{tabular}{|c |c || r | r | r |}
\cline{3-5}
\multicolumn{1}{c}{} & \multicolumn{1}{c}{}  &  \multicolumn{3}{|c|}{Geonames}  \\
\hline
Category & Pattern &  M$K^2$-tree &  I$K^2$-tree & RDF-3X \\
\hline
&  (S,P,O) & \textbf{1.8}  & 3.9 & 2,346.5 	\\
& (S,P,?) & \textbf{64.9} & 110.4 & 4,882.3 \\
& (?,P,O) & \textbf{0.1} &  0.3 & 0.6  \\
\multirow{-4}{*}{\makecell[c]{Bounded \\ Predicate}} & (?,P,?) &\textbf{0.4} &  0.5 & 0.7 \\
\hline
& (S,?,O) & 5.3  & \textbf{4.4} & 6,118.6 \\
& (S,?,?) & 95.0 & \textbf{69.7} & 229.7 \\
\multirow{-3}{*}{\makecell[c]{Unbounded \\Predicate}} & (?,?,O) & 240.0 & \textbf{187.0} & 2,473.1 \\
\hline
\multicolumn{5}{c}{} 
\end{tabular}

\begin{tabular}{|c |c || r | r | r |}
\cline{3-5}
\multicolumn{1}{c}{} & \multicolumn{1}{c}{}  &  \multicolumn{3}{|c|}{DBpedia}  \\
\hline
Category & Pattern &  M$K^2$-tree &  I$K^2$-tree & RDF-3X \\
\hline
&  (S,P,O) & \textbf{3.2} &  6.2 &2,532.4  \\
& (S,P,?) & \textbf{358.7} &   608.5 &4,117.3\\
& (?,P,O) & \textbf{0.6} &  1.6 &143.9\\
\multirow{-4}{*}{\makecell[c]{Bounded \\ Predicate}}& (?,P,?) & \textbf{ 0.7} &  1.6 &0.9\\
\hline
& (S,?,O) &  7,186.1 & \textbf{155.2} &6,330.6\\
& (S,?,?) &  3,925.2 & 911.2 & \textbf{272.3}\\
\multirow{-3}{*}{\makecell[c]{Unbounded \\Predicate}} & (?,?,O) &  10,918.1 &  1,444.6 & \textbf{1,377.9}\\
\hline
\end{tabular}

\caption{Time evaluation of simple patterns for RDF, in $\mu$s per result}
\label{tab:tempRDF1}
\end{table}

For bounded-predicate queries, the multiple $K^2$-tree is the fastest representation, and the query times of the \iktree are about twice those of the multiple \ktree. This is consistent with the expected results: in query patterns with bounded predicates, the query can be answered simply performing a row, \mbox{column} or full-range query over a single predicate, and therefore only a single \ktree must be accessed in the multiple \ktree representation. In these queries, the \iktree is expected to obtain higher query times due to the increased complexity in the low-level navigation operations over the \iktree: it must execute additional \emph{rank} operations at each traversed node in order to count the number of ones in the current node bitmap and therefore obtain the number of bits in its children. In general, RDF-3X is much slower than the other approaches.

For patterns with unbounded predicates, instead, the \iktree is able to clearly outperform the multiple \ktree representation. This difference is much more noticeable in the DBPedia dataset (lower sub-table), due to the much larger number of predicates, because if a multiple \ktree approach is used \emph{all} the \ktrees in the collection must be queried to retrieve their results; when using the \iktree representation, results are obtained much more efficiently in a single traversal of the data structure. In the Geonames dataset, representative of dataset with fewer predicates, the \iktree obtains still a significant improvement in this family of queries and obtains the best query times in all the query patterns. In the DBPedia dataset, even though the \iktree is much more efficient than the multiple \ktree, it is still slower than RDF-3X on the query patterns involving unbounded subjects or objects, even though it is relative close in query times to it. In the next section we will analyze the reasons for this and introduce an alternative query technique to improve efficiency of these queries in datasets with a larger partitioning variable such as DBPedia.

\section{A lazy evaluation}

The \iktree is designed to be much more efficient than a simpler approach based on a collection of independent \ktrees when answering queries involving the partitioning dimension in a ternary relation. Particularly, in the previous section we showed that its query times for RDF datasets were much better than a multiple \ktree in unbounded-predicate queries, especially when the number of different predicates (i.e. the size of the partitioning variable) is large. In spite of this improvement, the \iktree still failed to beat a state-of-the-art representation of RDF datasets in some unbounded-predicate query patterns.  

The navigation algorithms over the \iktree that we have introduced are still inefficient when performing unbounded-partitioning-variable queries: at each basic navigational step over the conceptual tree we still have to keep track of a list of active predicates that will be updated during traversal. To maintain and update this list, all the bits of each node must be checked, and the corresponding values for the next level are updated for each $one$ found. This process can be very expensive for nodes containing many predicates. Notice also that this process may be not only expensive but also unnecessary, since the complete branch may not yield any result when the traversal completes. Therefore, all this greedy process of computing, for each level, which value of $Y$ corresponds with each bit of the current node could be avoided for many branches that do not produce any result.

In this section we  propose a \emph{lazy evaluation} strategy for the navigation of the conceptual \iktree. This strategy delays the computation of real values, storing only the minimum amount of information during a top-down active values of the predicate in each node until a result is found in the branch. This navigation strategy is designed to optimize the performance of queries with unbounded partitioning variable in datasets with a large partitioning dimension. 

The lazy evaluation approach subdivides the search process in two steps: first, a top-down traversal of the tree retrieves the results of the query and then a bottom-up traversal of the tree is performed to recover the $Y$ values for each result when necessary. The main difference in the top-down traversal of the tree is that the list of active values is not computed at each node. Instead of computing the complete list, the only value that is computed is the number of bits set to $one$, like in a fixed-partitioning-variable query. This count of the number of $ones$ in a node $N_i$ can be easily performed with additional rank operations to obtain $rank_1(N_j)=rank_1(T,init+|N|)-rank_1(T,init)$, where $init$ is the position of the first bit of $N_i$ in the tree ($T$). Hence, the top-down traversal only takes into account the number of values of $Y$ that are active in each node, but not the actual values of $Y$ associated with each bit.

Once the leaves of the conceptual tree are reached, the real values of $Y$ must be computed in a bottom-up traversal of the tree. The main advantage of this technique is that this additional computation is only performed for the branches that have actually led to a valid result. This bottom-up traversal starting at a leaf node performs the inverse mapping of the one computed in the eager evaluation: at each node, the list of \textit{relative} values within that node is computed. In that way, a bit $one$ in the position $m$ of a leaf node has initially associated the value $y_m$; this value will be mapped in the upper levels of the tree, updating it in each step of the bottom-up traversal until its real value is obtained at the root of the tree. Given a node with a list of \textit{relative} values, they are transformed by checking the bitmap of the parent node. A \textit{relative} value $y_j$ in a node will be mapped to the position of the $j$-th $one$ in the bitmap of the parent node. The process continues by recursively updating the lists of \textit{relative} values, mapping each value with its position in the parent, until the root is reached. When the root is reached, the actual position of the node determines the final value of $Y$ corresponding to each item. The selection of the $j$-th one in the bitmap of the parent node can be efficiently computed with a $select_1$ operation, that reverses the $rank$ operations performed in the greedy computations.  

An additional optimization of this process can be performed for such queries where dimensions $X$, $Z$ or both are unbounded, or when a range of values is queried in $X$ or $Z$. In those query patterns, several branches are explored, so many results could have common ancestor nodes, that have to be explored many times (once for the mapping of relative values of each result). When many branches are expanded in a query, a \textit{merge sort} of the relative lists for all the children of a node can be performed, obtaining a single list of relative values which is mapped once using the bitmap of the parent node. In this way, several lists of results can be kept at each step of the traversal, all of them sorted by their $Y$ value; the lists of sibling nodes are progressively merged during the bottom-up traversal before updating them with the contents of the parent node, hence avoiding redundant mapping operations.

\begin{figure}
  \centering
      \includegraphics[width=0.7\textwidth]{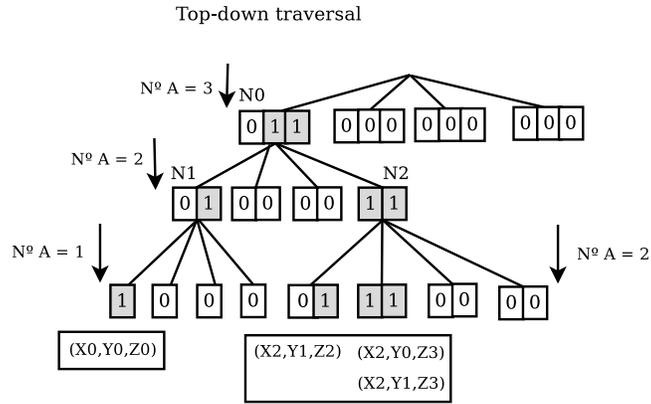}      
  \caption{First step of the top-down traversal of the lazy evaluation}
  \label{fig:IKLazy1}
\end{figure}

\begin{figure}
  \centering
      \includegraphics[width=0.75\textwidth]{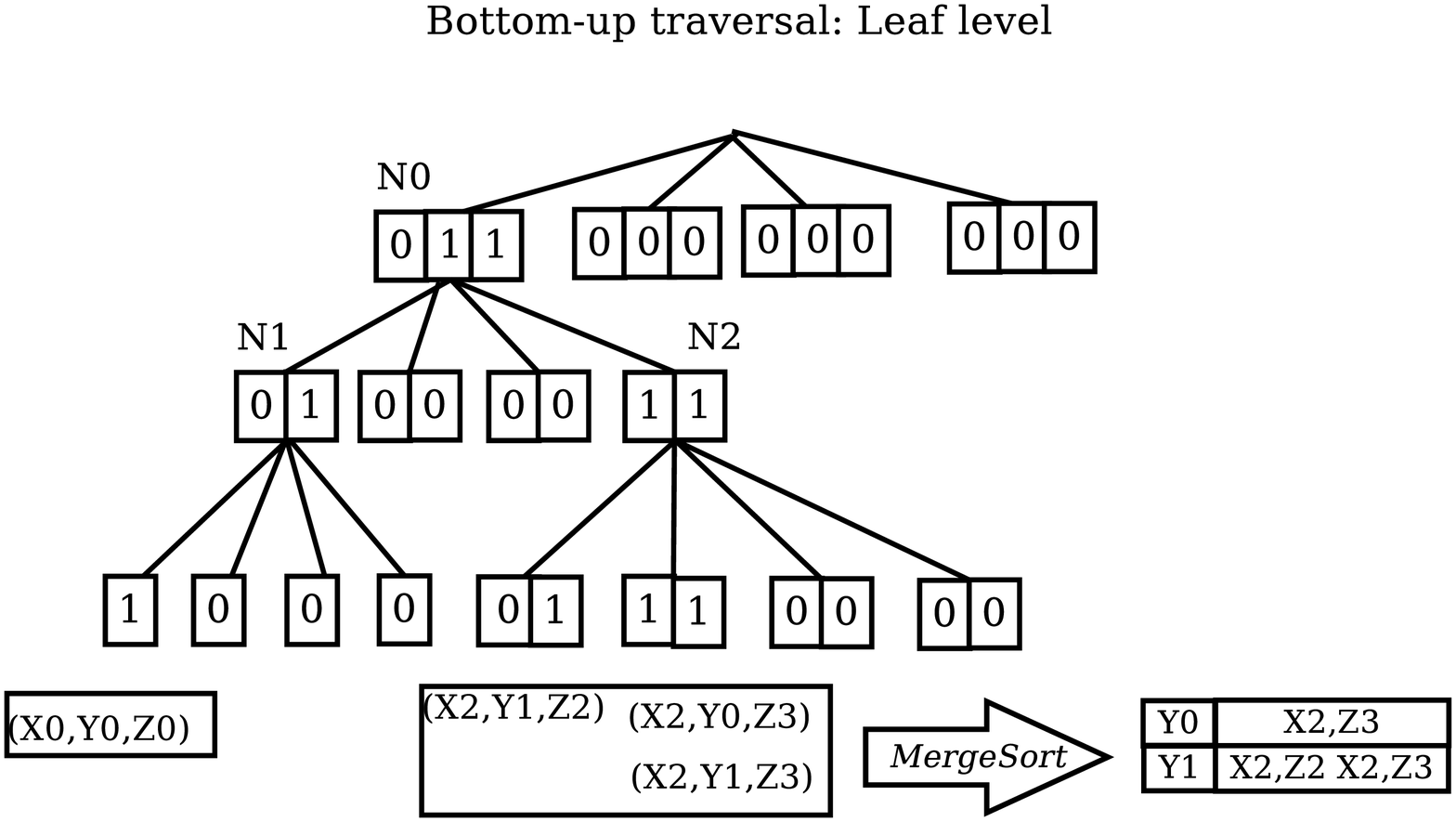}
  \caption{Reaching the leaf level of the lazy evaluation}
  \label{fig:IKLazy2}
\end{figure}

\begin{figure}
  \centering
      \includegraphics[width=0.85\textwidth]{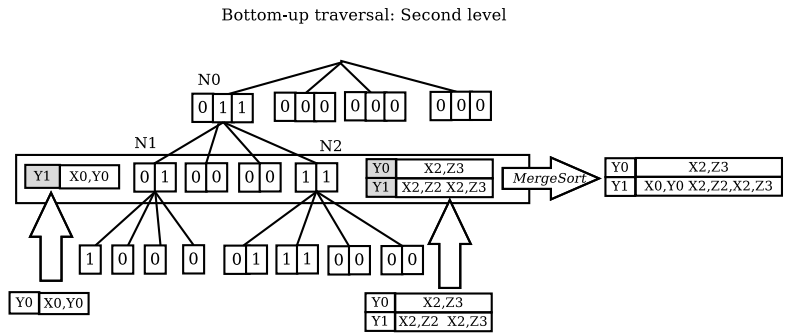}
  \caption{Starting the bottom-up traversal of the lazy evaluation}
  \label{fig:IKLazy3}
\end{figure}

\begin{figure}
  \centering
      \includegraphics[width=0.75\textwidth]{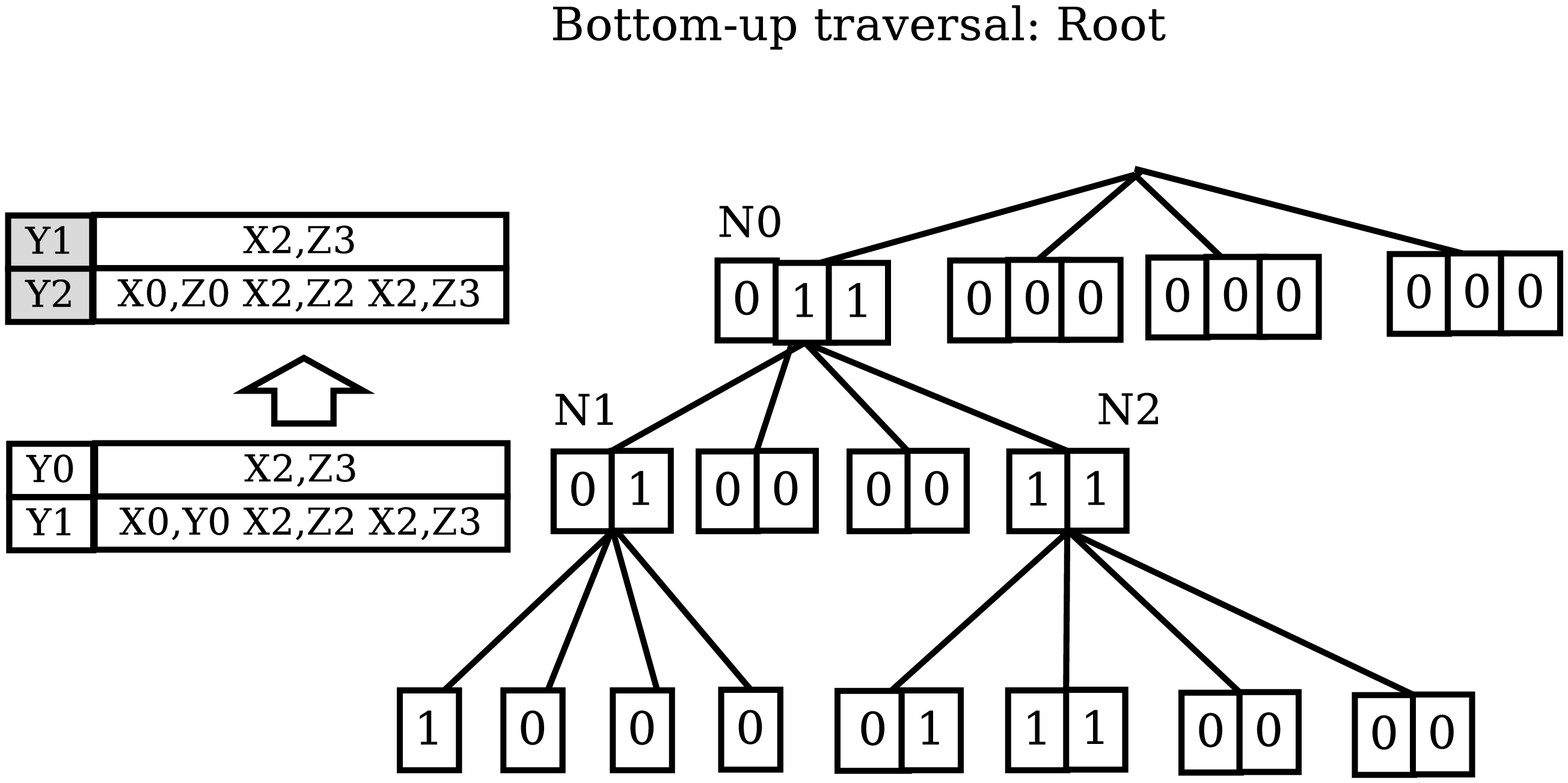}
  \caption{Reaching the root in lazy evaluation}
  \label{fig:IKLazy4}
\end{figure}

To illustrate the lazy strategy we present a complete example of the query algorithms using lazy evaluation for a dataset with three different values for the variable $Y$ through a set of figures. Figure \ref{fig:IKLazy1} shows the initial top-down traversal performed over the tree. For each node, only the number of active values are stored (in order to know the size of the children). In this way, node $N_0$ contains 2 bits with $one$ value so its children has 2 bits each one. $N_1$ contains one bit with $one$ value so its leaves only contain one bit. On the other hand, the two bits with value $one$ of $N_2$ produce leaves with two bits each one. At the end of this process, four results are obtained, which are highlighted in the figure. Their values for the variable $X$ and $Z$ are given by the branch of the tree which they are located. Their values for the $Y$ variable are unknown and a \textit{relative} value is computed as their position in the leaf node they belong. For instance, the leaf $(x_2,y_1,z_2)$ is related to the second value of $Y$ because is located in the second position of the leaf node. In this point the down-top traversal starts, shown in Figure \ref{fig:IKLazy2}. First, a merge sort is performed to join the lists of results which belong to the same parent node. The element $(x_0,y_0,z_0)$ is in an independent branch. However, the other three results $\{(x_2,y_1,z_2),(x_2,y_0,z_3),(x_2,y_1,z_3)\}$ are merged in the same list: $y_0$ has one result associated $(x_2,z_3)$ and $y_1$ contains two results $(x_2,z_2),(x_2,z_3)$. Next step consists of transforming the current relative values in basis to the parent nodes. Then, as the Figure \ref{fig:IKLazy3} shows, the list $y_0$ with the element $(x_0,z_0)$ is transformed to $y_1$ since the first bit one in the parent $N_1$ is in the second position. However, the list $y_0:(x_2,z_3)$ continues associated to the $y_0$ value since the first one in the parent $N_2$ is in the first position. The same happens with $y_1:(x_2,z_2),(x_2,z_3)$ which is transformed to $y_1$ because the second one of the parent node $N_2$ is in the second position. The three lists of this level share the same parent node so they are merged by $Y$ value. Finally, in Figure \ref{fig:IKLazy4}, the lists are transformed with the information of $N_0$. The first one of $N_0$ is in the second position and the second one is in the third position, so the final results will be $(x_2,y_1,z_3),(x_0,y_2,z_0), (x_2,y_2,z_2),(x_2,y_2,z_3)$. 

\subsection{Evaluation on RDF datasets}

To measure the efficiency of the lazy evaluation strategy we compare it with the basic navigation strategy in the RDF dataset DBPedia, the dataset with a large number of predicates. We focus only on the query patterns that involve an unbounded partitioning variable (unbounded predicate), and particularly on the query patterns $(S,?,?)$ and $(?,?,O)$, in which many branches of the conceptual tree were traversed and the \iktree was slower than RDF-3X in our previous experimentation. We do not show the results for the query pattern $(S,?,O)$ because in this query pattern only a single branch of the conceptual tree is expanded; hence the advantages of these technique become negligible and the additional navigation cost of the lazy evaluation would dominate the overall query time. The same experimental setup and query sets were used in these experiments, adding the lazy \iktree algorithms to the results.

\begin{table}
\centering
\begin{tabular}{|c | r | r | r | r | r|}
\hline
Query &  M $K^2$-tree &	Interleaved $K^2$-tree &  Lazy I$K^2$-tree & RDF-3X \\
\hline
(S,*,*) & 3,925.2	& 911.2 & \textbf{232.7} & 272.3\\
(*,*,O) & 10,918.1 	&1,444.6	&	\textbf{430.8} & 1,377.9\\
\hline
\end{tabular}
\caption{Time, in $\mu$s per result, of basic and lazy evaluation in patterns with unbounded predicate on DBpedia}
\label{tab:lazy}
\end{table}

Table \ref{tab:lazy} shows the results obtained. The lazy evaluation improves significantly the results for the studied query patterns, and achieves query times up to 5 times faster than the original eager algorithm, becoming the fastest representation in both query patterns. This new evaluation technique, combined with the original eager algorithms, make the \iktree faster than RDF-3X in all the unbounded-predicate queries in our experiments, therefore making the \iktree the most consistent alternative in space and query times. Note also that the application of a lazy evaluation technique can be easily determined depending on the characteristics of the dataset and the query; hence simple heuristics can be used to determine the appropriate evaluation strategy for each case.  
 

\section{Differential \iktree}

In the previous sections we introduced the \iktree as an alternative for the representation of a collection of binary relations, showing that it is much more efficient than the equivalent collection of \ktrees when many of these \ktrees would be accessed in a single query. In this section we explore an alternative application of the \iktree to the representation of binary relations and their changes, that is, the evolution of binary relations along time. Our proposal is a ``differential'' representation of the data that attempts to achieve maximum compression representing only the changes in the data instead of a complete snapshot of the data at each time instant.

Let us consider a temporal graph that models the evolution of a binary relation over time, that is, a ternary relation over $X \times Y \times T$, where $X$ and $Y$ are sets of nodes (and usually $X = Y$) and $T$ is a set of time instants at which a relation changed (that is, an edge in the graph appeared or disappeared). Notice that, when following a vertical partitioning approach, we will consider in this domain that time ($T$) is our partitioning variable, and $X$ and $Y$ are our regular dimensions.

The operations of interest in this domain involve usual queries on graph nodes, that is, direct or reverse neighbors of specific nodes, or checking whether there was a connection between two specific nodes. These basic graph queries will be increased in a temporal graph with a temporal constraint. \emph{Time-instant} queries refer to a specific time instant $t$ in $T$, and ask for the state of the graph at that specific $t$. \emph{Time-interval} queries refer to one or more intervals in time, which were the neighbors of node $x$ between times $t_5$ and $t_{13}$. Time-interval queries can have different semantics that determine the expected results from a query: \emph{weak} time-interval queries ask for the relations that occurred at \emph{any} point during the interval, whereas \emph{strong} time-interval queries would only return the relations that were active at \emph{all} points of the time interval.

In \cite{dcc} several proposals based on collections of \ktrees were introduced. These proposals based on \ktrees allow simple query in simple binary relations, but querying a collection of \ktrees used to represent a temporal graph proved to be very inefficient in time-interval queries, due to the need to synchronize access to an arbitrary number of \ktrees in time-interval queries, leading to poor performance as the time dimension increased in size, and being overcome by simpler strategies based on the pure compression of log changes per node.

A first approach to represent a temporal graph using \ktrees would store independent \ktrees for each time instant, each storing a complete snapshot of the graph. This approach essentially uses $T$ (time) as a partitioning variable and stores the state of the graph at each time instant. This approach relies on the compression capabilities of the \ktree to efficiently store the data, but it is unfeasible in domains with very large graphs and relatively few changes. An \iktree representation of the same snapshots would be more efficient in queries but still too large. 

The \iktree, based on vertical partitioning but providing efficient access to ranges in the partitioning dimension, is well-suited to represent time-evolving binary relations in a ``differential'' way. Its ability to represent a collection of \ktrees in a single data structure allows it to follow a space-efficient philosophy, storing only changes in the graph, and its indexing capabilities in the time dimension aim to provide query efficiency close to that of a naive snapshot-based representation. To demonstrate the efficiency of the \iktree to handle ternary relations in comparison with previous, simpler approaches based on collections of independent \ktrees, we present here an experimental comparison between both alternatives for the compact representation of temporal graphs. 

Let us consider a representation of a time-evolving graph based on a collection of \ktrees. In our proposal, we will represent each time instant (except $t_0$) as a change log. That is, at $t_0$ we store a complete snapshot of the graph (all triples $(x,y,t_0)$ in the ternary relation). At $t_k$, for $k >
0$, we store $(x_i,y_j,t_k)$ in tree $k$ iff the relation between $x_i$ and $y_j$ changed between $t_{k-1}$ and $t_k$, that is, we store for each edge of the graph the time instants when the edge appears or disappears.  Notice that, because of how \ktrees are built, in each differential \ktree an internal node in a \ktree $K_i$ will be set to 1 if and only if there has been at least one change in the region it covers between $t_{i-1}$ and $t_i$.

\begin{figure}[ht!]
\centering
  \includegraphics[width=\textwidth]{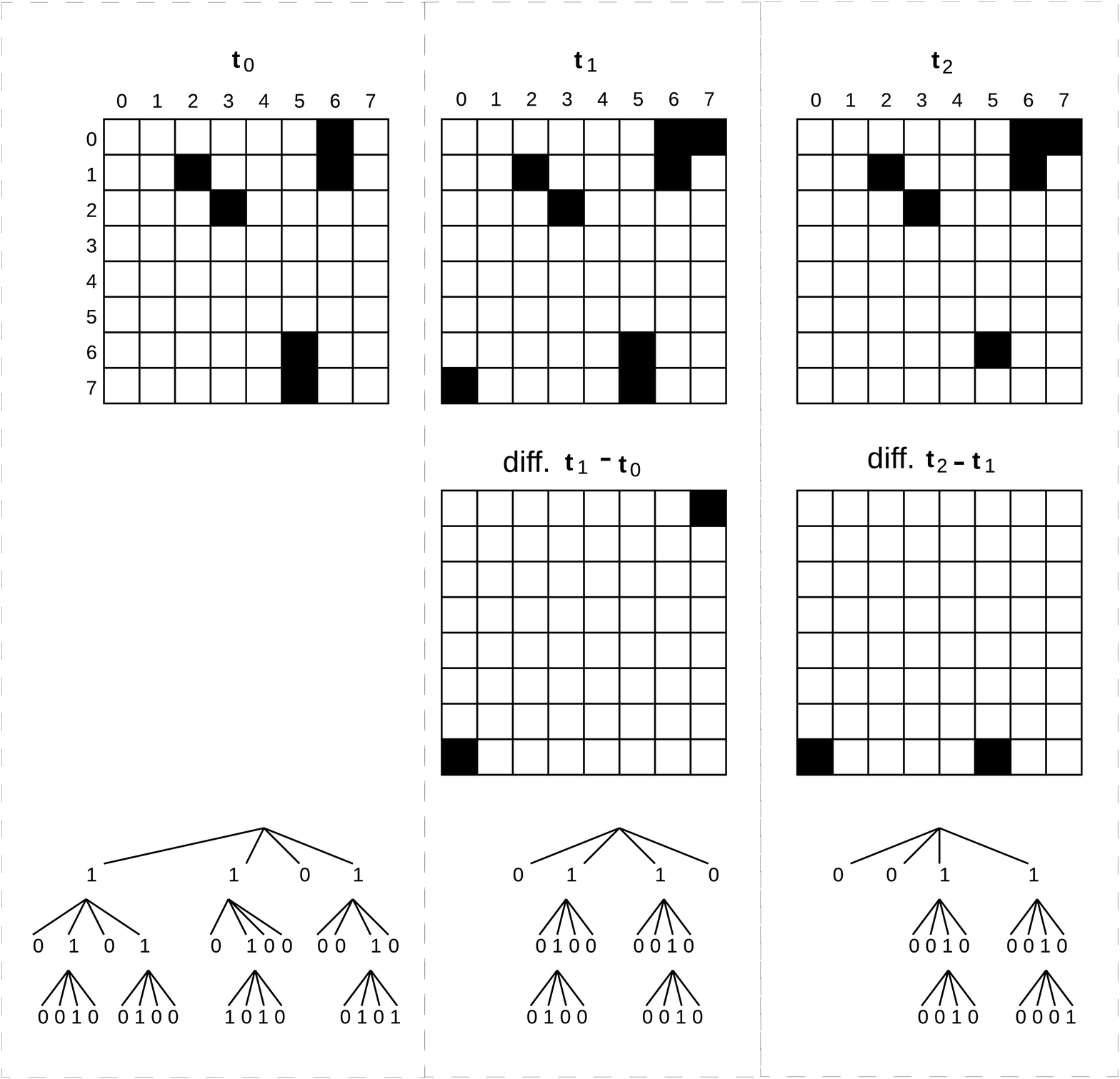}
  \caption{Representation of a temporal graph with ``differential'' \ktrees.}
\label{fig:interleaved:temporalktree}
\end{figure}

Our representation can be seen as a ``differential'' approach that represents the complete state of the graph at the beginning but only the changing occurring in the successive time instants. Considering a typical domain, where the number of changes in a graph at any time instant is much smaller than the overall number of edges, the expected space results would be much smaller than those of a naive snapshot-based approach. Figure \ref{fig:interleaved:temporalktree} shows a ``differential'' representation of a graph using a collection of differential snapshots, each encoded using a simple \ktree. The original full snapshots for each time instant are shown at the top (black cells represent items in the adjacency matrix); in the middle we show the corresponding ``differential'' snapshots for the time instants 1 and 2, and finally the \ktree representations for the full snapshot at $t_0$ and the differential snapshots at $t_1$ and $t_2$ are shown at the bottom.

Despite its increased compression, if we use a simple collection of \ktrees to store the snapshots, the query efficiency of this proposal is severely limited by the efficiency to solve constraints in the time dimension. Using this representation, a relation exists between $x_i$ and $y_j$ at time $t_k$ iff there is an odd number of triples $(x_i,y_j,t_m)$, where $m \in [0..k]$. Therefore, any query involving a time instant must be able to efficiently query the representation of the graph at all times before that time in order to count the number of changes. This count operation can be implemented with an x-or operation involving all the \ktrees corresponding to the interval $[t_0, t_k]$, that requires a synchronized traversal of all the trees in the interval. To answer time-interval queries $(x,y, I=[t_{\ell},t_r])$ we need to perform a similar operation adapted to the semantics of the query. To answer \emph{weak} time-interval queries semantics, that return all elements active at any point in the interval, a cell will be returned if it was active at $t_{\ell}$ or if it changed its state at any point between $t_{\ell}$ and $t_r$. We can compute this performing an x-or operation to check if the cell was active at $t_{\ell}$ and an or operation in the interval $[t_{\ell}, t_r]$ to check for changes within the query interval. Again, this operation must be checked for each node expanded in the \ktrees depending on the range $R$ determined in the other dimensions. Operations are similar for strong time-interval queries, required a synchronized traversal of multiple \ktrees performing logical operations on the collection: in \emph{strong} queries we return only elements active during the complete interval, so we must check that the element was active at $t_{\ell}$ and that no changes occurred in $[t_{\ell}, t_r]$.

Therefore, using a simple approach based on multiple \ktrees, a purely differential approach like the one just proposed faces the usual limitations in vertical partitioning: queries involving the partitioning variable, time, may have poor efficiency. In addition, the proposed representation requires any query in the temporal graph to access a variable number of \ktrees, and that number depends not on the length of a query interval but simply on the time instant of the query.

A proposal based on the \iktree for the same representation requires essentially the same space, as explained previously. The \iktree can be seen as a combination of the multiple \ktrees in a single data structures, combining their corresponding nodes in the final conceptual tree, so that their bits are placed together in the final physical representation and can be accessed more efficiently. This makes the \iktree able to allow simpler access to a sequence of values in the third dimension. In our case, this means that once a single leaf node in the tree is accessed (an edge in the graph), all the changes for that edge will be stored consecutively and the complex synchronized traversals of multiple \ktrees are reduced to much simpler counting operations in the \iktree bitmap.

\begin{figure}[ht!]
\centering
  \includegraphics[width=0.8\textwidth]{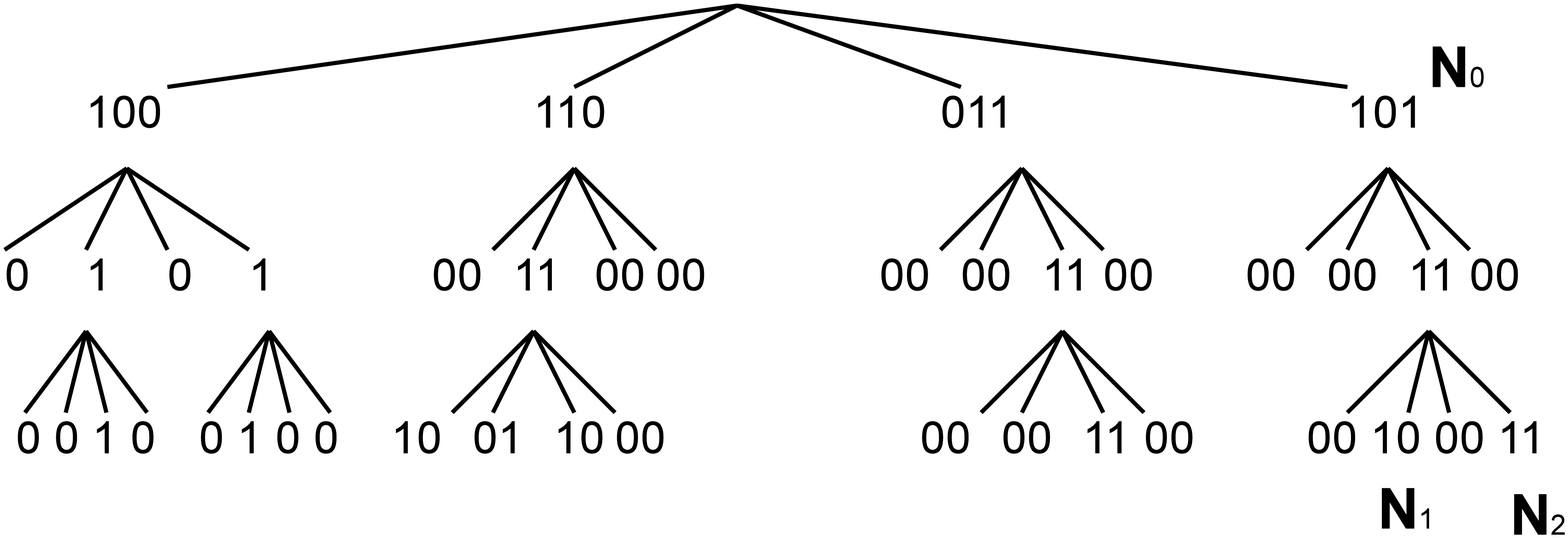}
\caption{``Differential'' \iktree representation of the graph in Figure \ref{fig:interleaved:temporalktree}}
\label{fig:interleaved:temporaldiffiktree}
\end{figure}

 Figure \ref{fig:interleaved:temporaldiffiktree} shows the representation of the same graph in Figure \ref{fig:interleaved:temporalktree} using a ``differential'' \iktree representation. In each leaf of the \iktree we store a bit for each change in the associated edge. In the internal nodes of the \iktree we store a bit for each time instant where at least one cell covered by that node suffered a change. For example, node $N_1$ represents row 6, column 5 of the adjacency matrix. This cell is active at $t_0$ and does not change at any other instant; hence its bitmap contains a single 1 in the first position, corresponding to $t_0$ (the actual mappings of the bits are $t_0$ and $t_2$, as we can see in the bitmap of $N_0$). On the other hand, node $N_2$, that represents row 7, column 5, is active at $t_0$ and becomes inactive at $t_2$; therefore its bitmap contains two bits set to 1, one for each change. Notice that in this case the \iktree representation provides a huge advantage to perform the counting operations required by queries, since the changes for each node are consecutive in the final bitmap representation. As we will see, all queries can be rewritten in terms of counting operations that are solved by means of rank operations in the bitmaps of the \ktree:

\begin{itemize}
  \item To answer a time-instant query, $(x, y, t_i)$, we navigate the \iktree like in a fixed-value query. In each internal node, we only stop navigation in the branch if the bitmap of the current node has all bits set to 0 until (and including) the bit corresponding to $t_i$: this case indicates that all cells in the current submatrix were 0 at $t_0$ and never changed. When we reach a leaf node, we can know whether the cell was active at $t_i$ counting the number of ones between $t_0$ and $t_i$, which can be easily computed with 2 rank operations on the bitmap $L$. Notice that we need to add rank support to $L$, which is not needed in basic \ktrees, but the small increase in space provides a much more efficient query method. The cost of a time-instant query does not depend anymore on $t_i$, the number of time instants before the current one. However, in order to provide efficient $rank$ support on $L$ we may disregard some of the improvements to compression that could be achieved using for instance a statistical compression of the last levels using DAC. 
  \item To answer time-interval queries, we behave exactly like in time-slice queries: we navigate until we reach the leaves, and in the leaves we check the number of changes using just rank operations. As explained, the operations to obtain results depend on the semantics: in \emph{weak} queries we need to count the number of changes up to the beginning of the interval (i.e. check if the relation was active at the beginning) or whether a change occurred in the interval (i.e. the relation appeared at some point in the interval). On \emph{strong} queries, we need to make sure that the relation was active at the beginning of the interval and no changes occurred in the complete interval. In both cases, operations are reduced to simple $rank$ operations to count $ones$ in the bitmap of the leaf node.
\end{itemize}

\subsection{Experimental comparison with independent \ktrees}

To demonstrate the indexing capabilities of the \iktree for the representation of the temporal evolution of graphs or binary relations, we compare it with a simple collection of ``differential'' \ktrees for the representation of different real and synthetic time-evolving graphs. 

We analyze the temporal evolution of three different graphs: the \emph{CommNet} dataset is a synthetic graph simulating a small, highly active network where relations are created randomly and last for a few time instants; the \emph{Monkey Contact} dataset is a collection of snapshots taken from a real social network; the \emph{Power Law} dataset is also a synthetic graph where edges follow a power-law degree distribution, where the first nodes in numbering order contain much more relations than the others. The three datasets are represented using our \iktree and also using multiple \ktrees that encode the same change logs (M\ktree). 

\begin{table}[ht!]
\centering
	\begin{tabular}{|c|r|r|r|r|r|r|}
	\hline
	Collection     & Size & Snapshots   &	change	& nodes/ 	& edges/   \\
				   &   (MB)		&			  &	rate	& snapshot	& snapshot      \\
	\hline
	CommNet		   &   225.9	& 10,000	  &	25\%	&	  10,000	&	  20,000	\\
	Monkey Contact &  4,303.9	& 220	 	  &	1\% 	&  3,200,000	&   2,500,000   \\
	Power Law	   & 22,377.1	& 1,000 	  &	2\%		&  1,000,000    &   2,900,000	\\
	\hline
	\end{tabular}
	
	\caption{Temporal graphs used in our experiments.}
	\label{temporaldatasets}

\end{table}

Table \ref{temporaldatasets} summarizes some basic information of the datasets. It aims to give a raw estimation of the size of the dataset. The first column shows a simple estimation of the ``base size'' of the dataset, given by the size of all the edges in all of the snapshots of the graph, each stored as a pair of integers. The next columns represent the number of snapshots (time instants) considered in the sequence, the change rate (percentage of edges that appear/disappear at each time instant, over the average number of edges) and finally the (average) number of nodes and active edges per snapshot of the graph.

Table \ref{tab:spT2} shows a comparison in space between the \iktree and a collection of independent \ktrees. As reference and baseline for comparison, we also include the number of snapshots in each dataset and an additional column with the ``differential size'' of the dataset. This size is the space of a plain representation of the first complete snapshot (as pairs of integers) and a representation of each change as a tuple (source, destination, appear/disappear); a smaller differential size indicates a graph with few nodes and/or few changes, and acts as a simple baseline for the space efficiency of a compact representation based on encoding lists of differences. 

For the comparison we build ``equivalent'' versions of the M\ktree and the \iktree, using the same values of $K$ ($K=4$ in the first level and $K=2$ in the remaining levels), without applying a statistical compression in the lower levels. In our results the \iktree is even smaller than a collection of independent \ktrees, due to the fact that we use an unmodified \ktree structure that stores some internal information about the size of the matrix and additional parameters of the internal tree structure; this information becomes redundant in this case, since we have a collection of \ktrees representing graphs of the same size and with the same internal structure. If the redundant information in the \ktrees were omitted, the \iktree would be slightly larger than the individual \ktrees, since the only real difference in the size of the data structures would be due to the addition of a rank structure to the last level of the conceptual tree, in order to perform efficiently counting operations. 

\begin{table}
\centering
\begin{tabular}{|c | r | r | r | r | r|}
\hline
Dataset & $|Snapshots|$ & ``Diff. size'' & M\ktrees &  \iktree \\
\hline
\hline
CommNet  & 10.000 &	716.4		& 137,28 & 136,51\\
\hline
Monkey contact & 220 &	222	& 281,11 & 281,03\\
\hline
Power Law  & 1.000 & 33.5	& 4,55 & 4,54\\

\hline
\end{tabular}
\caption{Space comparison on different temporal graphs (in MB)}
\label{tab:spT2}
\end{table}

\begin{table}
\begin{tabular}{|l | r | r || r | r || r | r |}
\hline
     &  \multicolumn{2}{c||}{CommNet}  & \multicolumn{2}{c||}{Monkey Contact} & \multicolumn{2}{c|}{Power Law}\\
\hline
   Patterns & M$K^2$-tr &  I$K^2$-tr&  M$K^2$-tree &  I$K^2$-tree & M$K^2$-tr &  I$K^2$-tree\\
\hline
Instant & 86,78 & \textbf{1,7} & 0,27 & \textbf{0,14} & 17,60 & \textbf{1,28}\\
\hline
Weak  & 87,19 & \textbf{1,84} & 0,27 & \textbf{0,14} & 20,72 & \textbf{1,56}\\
\hline
Strong  & 88,02 & \textbf{1,82} & 0,26 & \textbf{0,16} & 19,73 & \textbf{1,44}\\
\hline
\end{tabular}
\caption{Time comparison: direct neighbors of a node on temporal graphs (in ms/query)}
\label{tab:tempTemp}
\end{table}

\begin{table}
\begin{tabular}{|l | r | r || r | r || r | r |}
\hline
     &  \multicolumn{2}{c||}{CommNet}  & \multicolumn{2}{c||}{Monkey Contact} & \multicolumn{2}{c|}{Power Law}\\
\hline
   Patterns & M$K^2$-tr &  I$K^2$-tr&  M$K^2$-tree &  I$K^2$-tree & M$K^2$-tr &  I$K^2$-tree\\
\hline
Instant & 89,07 & \textbf{1,79} & 0,28 & \textbf{0,1}5 & 19,86 & \textbf{1,43}\\
\hline
Weak  & 90,96 & \textbf{2,07} & 0,26 & \textbf{0,15} & 18,36 & \textbf{1,46}\\
\hline
Strong  & 93,75 & \textbf{2,05} & 0,26 & \textbf{0,16} & 19,98 & \textbf{1,50}\\
\hline
\end{tabular}
\caption{Time comparison: reverse neighbors of a node on temporal graphs (in ms/query)}
\label{tab:tempTemp2}
\end{table}

Tables \ref{tab:tempTemp} and \ref{tab:tempTemp2} show the query performance achieved for the three datasets with the \iktree, in contrast with the multiple \ktree approach. We ask for direct and reverse neighbors of a given node at a time instant or time interval with strong and weak semantics. Each query set contains 2000 queries. Nodes and time instants are selected randomly on the size of the node set and the number of snapshots respectively. Different query sets are built for direct and reverse neighbors. Finally, for time-interval queries, we show the results for a fixed interval length of 100, hence limiting starting points of the interval so that the complete length of the interval will be valid. Note that the length of the interval has almost no effect on query times in the multiple \ktree representation when using a differential approach. In the \iktree query times do change with interval length, but the query cost is more related to the number of matched results than to an increased cost in querying for long intervals.

The experimental results show the superiority of the \iktree in all queries. The results are consistent in all the datasets and query types, with the \iktree being much faster to obtain the results. In the Monkey Contact dataset the difference is much smaller than in the other two datasets, due to the reduced number of snapshots existing in the collection (just over 200), which means that the number of instants that must be accessed are a significant percentage of the overall representation. As soon as the number of snapshots increases, the relative efficiency of the \iktree against a multiple \ktree approach also increases. This result is expected, since a purely differential representation in a collection of trees forces the query algorithms to traverse multiple trees simultaneously even to answer time-instant queries regarding the multiple \ktree approach in terms of querying efficiency in the context of temporal graphs. The huge improvement in the performance of time-specific constraints shows the relative efficiency of the \iktree against a simpler solution based on a simple collection of independent \ktree structures. 

Notice that in a ``differential'' representation the \iktree is much faster even in time-instant queries because all our queries (even time-instant queries) require traversing a series of time points to be answered. Particularly, in this differential approach the cost of the query depends more on the time point used in the query than on the length of the query interval: if the time-instant query refers to an early time point $t_i$ (or time interval $[t_\ell, t_r]$), we only need to check the changes in the interval $[0,t_i]$ (or $[0,t_r]$); on the other hand, a time-instant query that asks for a time point $t_j > t_i$ requires us to check more time instants and compute a higher number of changes. This is confirmed in our experiments, since time points and time intervals are selected randomly and the cost of time-instant and time-interval queries is very similar in Table \ref{tab:tempTemp} and Table \ref{tab:tempTemp2}, with only minor differences. The flexibility of the \iktree opens the possibility of exploring new representations such as this one, where instead of a simple vertical partitioning of the  datasets we perform an oriented, smarter partitioning focused on increasing compression or improving query support.

\section{Conclusions}

The \iktree is a compact data structure able to represent in very compact space ternary relations, providing indexing capabilities over the 3 dimensions and supporting different constraints in all of them with simple algorithms. The \iktree is especially well suited to ternary relations where one of the dimensions is smaller than the other two, since it follows a vertical partitioning approach that treats one of the dimensions in a different way. Nevertheless, the \iktree is able to answer queries involving fixed values or ranges in the partitioning dimension efficiently.

Our experimental evaluation shows that the \iktree is able to provide very compact representations in real-world datasets. In particular, our experiments show that a representation of RDF datasets using the \iktree improves on previous results based on \ktrees and is competitive with state-of-the-art proposals. 

We also introduce several variations of the \iktree that improve on the basic structure and increase its applications. A \emph{lazy evaluation} strategy is presented, that provides a more efficient query technique in datasets where the partitioning dimension is relatively large. A \emph{differential} \iktree representation is introduced to manage the state of a time-evolving binary relation, storing the changes at each time instant in a compact way while providing efficient methods to retrieve the actual status of the relation at any point in time.
\bibliographystyle{elsarticle-num}

\end{document}